\documentclass[aps,prl,amsmath,amssymb,superscriptaddress,showkeys,showpacs,twocolumn,floatfix,preprintnumbers]{revtex4-1}

\usepackage{graphicx}
\usepackage{subfigure}
\usepackage{hyperref}
\usepackage{ulem}

\usepackage{xcolor}
\usepackage[utf8]{inputenc}

\begin{document}

\preprint{MITP-23-049, ULB-TH/23-08, IFIC/23-41}

\title{Sphaleron freeze-in baryogenesis with gravitational waves from the QCD transition}

\author{Fei Gao}
\email{fei.gao@bit.edu.cn}
\affiliation{School of Physics, Beijing Institute of Technology, 100081 Beijing, China}
\author{Julia Harz}
\email[]{julia.harz@uni-mainz.de}
\affiliation{PRISMA+ Cluster of Excellence \& Mainz Institute for Theoretical Physics, Johannes Gutenberg-Universit\"at Mainz, 55099 Mainz, Germany}
\author{Chandan Hati}
\email[]{chandan@ific.uv.es}
\affiliation{Service de Physique Th\'eorique, Universit\'e Libre de Bruxelles, Boulevard du Triomphe, CP225, 1050 Brussels, Belgium}
\affiliation{Instituto de F\'{i}sica Corpuscular (IFIC), Universitat de Val\`encia-CSIC,\\C/ Catedratico Jose Beltran, 2, E-46980 Valencia, Spain}
\author{Yi Lu}
\email{qwertylou@pku.edu.cn}
\affiliation{Department of Physics and State Key Laboratory of Nuclear Physics and Technology,\\ Peking University, Beijing 100871, China}
\author{Isabel M. Oldengott}
\email[]{isabel.oldengott@uclouvain.be}
\affiliation{Centre for Cosmology, Particle Physics and Phenomenology, Universit\'e catholique de Louvain, Louvain-la-Neuve B-1348, Belgium}
\author{Graham White}
\email[]{g.a.white@soton.ac.uk}
\affiliation{School of Physics and Astronomy, University of Southampton,
	Southampton SO17 1BJ, United Kingdom}

\begin{abstract}
A large primordial lepton asymmetry is capable of explaining the baryon asymmetry of the Universe (BAU) through suppression of the electroweak sphaleron rates (``sphaleron freeze-in") which can lead to a first-order cosmic QCD transition with an observable gravitational wave (GW) signal. With next-to-leading order dimensional reduction and the exact 1-loop fluctuation determinant, we accurately compute the lepton asymmetry needed to realize this paradigm, finding it to be an order of magnitude smaller than previous estimates. Further, we apply an improved QCD equation of state capable of describing the phase transition line together with the critical endpoint leading to better agreement with lattice and functional QCD results. Based on this, we identify the range of lepton flavor asymmetries inducing a first-order cosmic QCD transition. We then extract the parameters relevant to the prediction of GW signal from a first-order cosmic QCD transition. Our result showcases the possibility of probing the sphaleron freeze-in paradigm as an explanation of BAU by future gravitational wave experiments like $\mu$Ares. 
\end{abstract}

\pacs{95.30.Tg, 
11.30.Fs, 
05.70.Jk, 
12.38.Lg 
}

\keywords{sphalerons, lepton asymmetry, cosmic QCD transition, gravitational waves}

\maketitle
\paragraph{Introduction---}
We know little about the Universe at temperatures before Big Bang Nucleosynthesis (BBN). Based on our knowledge of particle physics, we however expect that the first second of the Universe was filled with many exciting events, like the generation of a baryon asymmetry and a transition from a quark-gluon plasma to a hadronic phase. The new era of gravitational wave (GW) cosmology provides a window towards these early stages because, if the QCD transition was strongly first-order, this would leave a relic stochastic GW background visible today. The recent findings of several pulsar timing array collaborations \cite{NANOGrav:2023gor,NANOGrav:2023hde,EPTA:2023fyk,EPTA:2023sfo,EPTA:2023xxk,Reardon:2023gzh,Zic:2023gta,Reardon:2023zen,Xu:2023wog} have lead to an enormous increase of interest to this end.
On the other hand, the origin of the observed baryon asymmetry of the Universe (BAU) is one of the biggest open puzzles in particle physics and cosmology. According to the standard leptogenesis picture, an asymmetry is created in the leptonic sector and converted to the baryonic sector by sphaleron processes~\cite{Kuzmin:1985mm}. This would imply the lepton asymmetry to be of the same order of magnitude as the baryon asymmetry (i.e. tiny) - for a review see \cite{Davidson:2008bu,Chun:2017spz,Elor:2022hpa}. It was however shown that in the case of \textit{large primordial lepton asymmetries} the restoration of the electroweak symmetry gets prohibited, which can lead to the right amount of baryon asymmetry due to a suppressed sphaleron rate~\cite{Bajc:1997ky,McDonald:1999he,Bajc:1999he,McDonald:1999in,March-Russell:1999hpw,Barenboim:2017dfq}. We will refer to this mechanism of exponential suppression for the rate for sphalerons due to non-restoration of the electroweak symmetry (such that the conversion of lepton asymmetry into baryon asymmetry happens extremely slowly) as ``sphaleron freeze-in".
At the same time, large lepton asymmetries have an impact on the QCD epoch  \cite{Schwarz:2009ii,Wygas:2018otj,Middeldorf-Wygas:2020glx,Bodeker:2020stj,Vovchenko:2020crk,Hajkarim:2019csy} and it was shown very recently \cite{Gao:2021nwz} that they can render the cosmic QCD transition to be first-order. In this Letter, we present the novel possibility of testing the sphaleron freeze-in paradigm in the presence of large lepton asymmetries via the GW imprint from a first-order cosmic QCD transition. This is based on a careful study of the impact of large lepton asymmetries from temperatures above the electroweak scale down to the QCD epoch, improving upon existing literature in several ways: To study symmetry non-restoration, we apply dimensional reduction techniques based on~\cite{Kajantie:1995dw,Braaten:1995cm} to obtain the thermal potential at finite density. We find that the lepton asymmetries required to produce the observed baryon asymmetry are about an order of magnitude smaller than previous estimates based on a perturbative effective potential~\cite{Bajc:1997ky,Barenboim:2017dfq}.
Concerning the QCD epoch, the new  equation of state (EoS)~\cite{gao2023} is applied which improves over~\cite{Gao:2021nwz} by a better agreement with lattice QCD and an improved description of the thermodynamic variables around the critical endpoint (CEP). 
Building upon these, we show for the first time that experiments like $\mu$Ares~\cite{Sesana:2019vho} can potentially probe the sphaleron freeze-in paradigm.

\paragraph{Constraints on lepton flavor asymmetries---}
We will denote the lepton flavor asymmetries as follows,
\begin{equation}
    Y_{L_{\alpha}} = \frac{n_{L_{\alpha}}}{s} = \frac{n_{\alpha}+n_{\nu_{\alpha}}}{s} \hspace{1cm}  (\alpha=e, \mu, \tau),
    \label{eq:lepton_asymmetry}
\end{equation}
where $n_i$ are particle minus anti-particle number densities, $s$ is the entropy density of the Universe, and the total lepton asymmetry is given by $Y_L=\sum_{\alpha} Y_{L_\alpha}$. 
Observational bounds on the lepton flavor asymmetries are many orders of magnitude weaker than on their baryonic counterpart. They are based on analyses of data from the anisotropy spectrum of the Cosmic Microwave Background (CMB) and measurements of primordial element abundances as predicted by Big Bang Nucleosynthesis (BBN). 
As it was shown in \cite{Dolgov:2002ab,Wong:2002fa} that neutrino oscillations tend to equilibrate lepton flavor asymmetries, it is usually assumed that during BBN and CMB the individual lepton flavor asymmetries $Y_{L_\alpha}$ are roughly one-third of the (conserved) total lepton asymmetry $Y_L$. Under this assumption, \cite{Oldengott:2017tzj} found $|Y_L| \leq 1.2\times 10^{-2}$ which is comparable to the constraints from BBN \cite{Pitrou:2018cgg}\footnote{Interestingly, recent measurements of the primordial helium abundance by the EMPRESS survey \cite{Matsumoto:2022tlr} could be interpreted as a hint towards a positive asymmetry in electron neutrinos \cite{Escudero:2022okz,Matsumoto:2022tlr}. }. Note however, that the level of equilibration depends on the details of neutrino oscillations \cite{Pastor:2008ti,Mangano:2010ei,Mangano:2011ip,Castorina:2012md,Barenboim:2016shh,Johns:2016enc} and the bounds of \cite{Oldengott:2017tzj} only hold approximately.
Finally, Ref.~\cite{Domcke:2022uue} recently presented constraints on the values of \textit{primordial} lepton flavor asymmetries (for $T>10^6$~GeV) using
 the hypermagnetic fields inducing overproduction of the baryon asymmetry. However, such a constraint is not applicable in our case because it requires the electroweak symmetry to be restored at high temperatures~\cite{Joyce:1997uy}.


\paragraph{Sphaleron freeze-in and baryogenesis---}
It is well known that the presence of a large background charge in the Universe can lead to non-restoration of symmetry at high temperatures. The various existing studies in the literature~\cite{Bajc:1997ky,McDonald:1999he,Bajc:1999he,McDonald:1999in,March-Russell:1999hpw,Barenboim:2017dfq} employ the perturbative effective potential at finite temperature and chemical potential. However, at finite temperature, the long-distance behavior becomes strongly coupled \cite{Linde:1980ts} and the hierarchy between some terms at different loop order breaks down (see \cite{Gould:2021oba,Curtin:2022ovx}). It becomes then necessary to re-organize perturbation theory in powers of the weak gauge coupling constant, $g$. Then one has control of defining the theory to an appropriate level of accuracy. This is automatically achieved using the framework of dimensional reduction ~\cite{Kajantie:1995dw,Braaten:1995cm}. At finite temperature, the imaginary time domain becomes a compact dimension of size $1/T$, resulting in a Kaluza Klein tower of heavy Matsubara modes. Integrating out such modes results in an effective three-dimensional theory where the hard modes screen the long-distance behavior and the theory becomes straightforward to organize in powers of $g$. The benefits of this approach can be quite drastic---GW predictions at one loop tend to differ from the predictions from a dimensionally reduced theory of $\mathcal{O}(g^4)$ by multiple orders of magnitude~\cite{Croon:2020cgk,Gould:2021oba}. Similarly, in our case, following Ref.~\cite{Gynther:2003za} we will define our theory again to $\mathcal{O}(g^4)$ and find a substantially different prediction of the needed lepton asymmetry compared to the existent results in the literature e.g. of~\cite{Barenboim:2017dfq,Bajc:1997ky}.
A detailed account of the full potential with all the technicalities will be presented in a longer and separate communication~\cite{Gao:2023lng}. However, at ${\cal O}(\mu^4/T^4,g^2 \mu^2 /T^2,g^3)$ an approximate form for the thermal potential is given by
\begin{eqnarray}
V_{\rm eff}(\phi) & \simeq & \frac{1}{2} \Big( -\frac{m_H^2}{2} + \frac{g^2 T^2}{16 m_W^2}(m_H^2+2m_W^2+m_Z^2+2m_t^2) \nonumber\\
&&- \frac{16}{121} \mu ^2  \Big) \phi ^2 
 - \frac{g^3 T}{32 \pi m_W^3} (2 m_W^3 + m_Z^3)\phi ^3\nonumber\\ &&+ \frac{1}{4} 
 \left( \lambda + \frac{9}{1331} \frac{\mu ^2}{T^2} \right)\phi^4 \,.
\end{eqnarray}
Note that the correction to the mass term, $-\mu ^2 \phi ^2$,  (which induces a negative correction to the quadratic) is responsible for the non-restoration of electroweak symmetry at high temperatures. The approach of dimensional reduction allows for numerical minimization of the effective 3-dimensional potential at a given finite temperature and chemical potential, which can be combined with an exact computation of the small-fluctuation determinant from e.g.~\cite{Carson:1990jm} to estimate the sphaleron freeze-in rate and the final baryon asymmetry in a reliable manner. 

\begin{figure}
    \centering
    \includegraphics[width=\columnwidth]{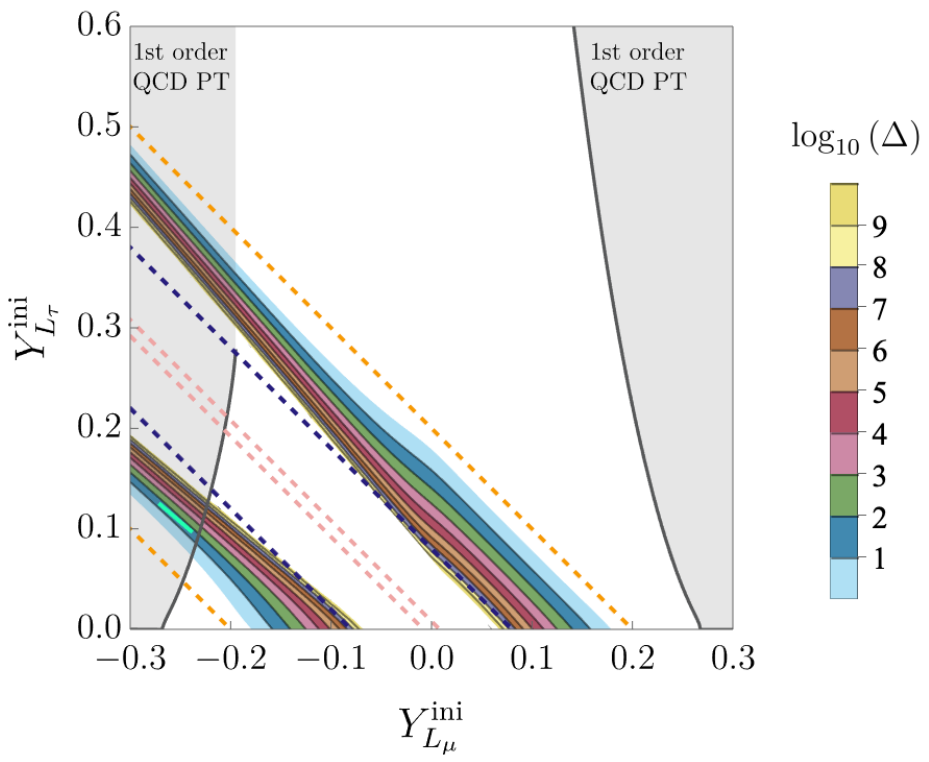}
    \caption{The viable parameter space for successful baryogenesis via sphaleron freeze-in in the plane of $Y_{L_\mu}^{\text{ini}}$ and $Y_{L_\tau}^{\text{ini}}$. The color coding of the bands corresponds to different values of entropy dilution (as shown in the legend) required to produce the observed baryon abundance today. The area within the pink, blue, and orange pair of dashed lines represents the CMB constraint in the limit of fully equilibrated lepton asymmetries~\cite{Oldengott:2017tzj}, corresponding to the cases $\Delta^{\text{FE}}=1$ (no entropy dilution), 10, and 25, respectively. The grey-shaded regions correspond to the parameter space that can induce a cosmic first-order QCD transition. The cyan line on the colored band corresponds to the benchmark line which we use to study the viability of probing GW signal from the first-order cosmic QCD phase transition (PT). Note that such a line leads to successful baryogenesis for an entropy dilution $\Delta\sim(18)$.}
    \label{fig:pot_genesis}
\end{figure}

Using this approach, we find the allowed initial lepton flavor asymmetries ($Y_{L_{\mu (\tau)}}^{\text {ini}}$) that can lead to the correct baryon asymmetry as shown in Fig. \ref{fig:pot_genesis}. Note that we choose $Y_{L_e}^{\text {ini}}$ to be vanishing without any loss of generality~\footnote{ An amount of $Y_{L_e}$ will be generated once the neutrino oscillations are in equilibrium shortly before BBN, however, it has no impact on our analysis. }. Interestingly, using our approach the prediction for primordial lepton asymmetry required for successful baryogenesis is reduced by an order of magnitude as compared to previous results in the literature, e.g. Refs.~\cite{Barenboim:2017dfq,Bajc:1997ky}. The colored contours show the allowed region capable of reproducing the correct observed baryon asymmetry. The primordial asymmetry $Y_{L_{\mu (\tau)}}^{\text {ini}}$ can get diluted before BBN in the presence of late-time entropy production. We define the entropy dilution factor as  $\Delta\equiv Y_{L}^{\text {ini}}/Y_{L}$. The differently colored contours correspond to different factors of entropy dilution required to produce the observed baryon asymmetry. We find that given the constraints from BBN and CMB on primordial lepton asymmetry (subject to flavor asymmetry equilibration due to neutrino oscillations), a minimum entropy dilution of $\Delta^{\text{NE}}_{\text{min}}\gtrsim 5$ (no-equilibration limit) and $\Delta^{\text{FE}}_{\text{min}}\gtrsim 18$ (full-equilibration limit) is required for reproducing the correct baryon symmetry. The entropy dilution can naturally occur via late-time decaying states in many common and well-motivated new physics scenarios e.g. long-lived moduli \cite{Moroi:1999zb}, supersymmetric condensate \cite{Thomas:1995ze}, gravitinos \cite{Moroi:1994rs},  inflatons \cite{Allahverdi:2002nb}, curvatons \cite{Moroi:2002rd}, dilatons \cite{Lahanas:2011tk}, $Q$-balls \cite{Fujii:2002kr}, etc. We now proceed to show that in the presence of entropy dilution as discussed above (after the QCD transition and before BBN), the primordial lepton asymmetry can lead to a predictive first-order QCD phase transition with an observable GW signal.

\paragraph{First-order QCD transition---} 
After the electroweak phase transition ($T \sim 100$~GeV) and before the onset of neutrino oscillations ($T\sim10$~MeV) the three lepton flavor numbers, the baryon number, and the electric charge of the Universe are conserved. In the absence of any entropy-producing process, the lepton asymmetries in Eq.~\eqref{eq:lepton_asymmetry} as well as the analogously defined baryon asymmetry $Y_B$ and the charge asymmetry $Y_Q$ are conserved quantities.
Each of these conserved charges is assigned a charge chemical potential. Assuming chemical equilibrium, the charge chemical potentials $\mu_B, \mu_Q, \mu_{L_{\alpha}}$ can be expressed through the chemical potentials of the individual particle species, e.g. $\mu_B = \mu_u + 2 \mu_d$ (where $(u,d)$  stands for the ($u,d)$-quark). Given the values of $Y^{\text{ini}}_B, Y^{\text{ini}}_{L_\alpha}$ and $Y^{\text{ini}}_Q$, the so called \textit{cosmic trajectory} is the solution for $(\mu_B, \mu_{L_\alpha}, \mu_Q)$ at a given temperature $T$. As pointed out in \cite{Schwarz:2009ii,Wygas:2018otj,Middeldorf-Wygas:2020glx}, large lepton flavor asymmetries shift the cosmic trajectory towards large quark chemical potentials (and hence a large baryon chemical potential). Ref.~\cite{Gao:2021nwz} applied results from functional QCD to express the thermodynamic quantities of QCD matter and found that sufficiently large $Y^{\text{ini}}_{L_\alpha}$ can indeed induce  
$\mu_{u,d}$ beyond the critical value at the $T_{\rm{CEP}}$, which implies a first-order QCD transition. The results of \cite{Gao:2021nwz} were based on solving the EoS using a gap equation from the Dyson-Schwinger equation for the propagator in the Rainbow-ladder (RL) truncation~\cite{Gao:2015kea}. 
At the time of the publication of Ref.~\cite{Gao:2021nwz}, this was the only \textit{QCD based} method delivering a complete computation of the phase diagram and the related thermodynamic quantities. However, as already pointed out in \cite{Gao:2021nwz}, the truncation scheme can be improved and in particular the location of the CEP from the RL truncation was not realistic. Therefore, by including the running of the quark-gluon vertex, the improved truncation gives the phase transition line (under the assumption of negligible surface tension) which agrees with lattice QCD results at low chemical potentials and makes predictions for the CEP location~\cite{Gao:2020qsj,Gao:2020fbl}.  The EoS of QCD can further be constructed from these results of functional QCD methods~\cite{gao2023}, giving the phase transition temperature at zero chemical potential at $T=157$~MeV, the curvature of the transition line $\kappa=0.016$ and a critical endpoint at ($T_{\rm{CEP}}$, $\mu_{\rm{CEP}}$) = (118, 200) MeV. The EoS has been incorporated in hydrodynamic simulations, allowing one to obtain particle yields and particle ratios that
are consistent with commonly used EoSs at high collision energies~\cite{gao2023}. We incorporate the improved thermodynamic quantities of ~\cite{gao2023} in the calculation of the cosmic trajectory. Compared to \cite{Gao:2021nwz}, the new method shows a significantly better agreement at low temperatures to the hadron resonance gas limit of \cite{Schwarz:2009ii,Wygas:2018otj,Middeldorf-Wygas:2020glx}. In this work, we are particularly interested in the values of $Y^{\text{ini}}_{L_\mu}$ and $Y^{\text{ini}}_{L_\tau}$ that induce a first-order cosmic QCD transition. As described in \cite{Gao:2021nwz}, imposing the condition
\begin{equation}
    |\mu_{u,d}| > \mu_{\text{CEP}} \hspace{1cm} \text{at T=$T_{\text{CEP}}$}
\end{equation}
additionally to the 5 conservation laws for the charge asymmetries allows us to find the values of $Y^{\text{ini}}_{L_{\mu}}$ (for given $Y^{\text{ini}}_{L_\tau}$ and $Y^{\text{ini}}_{L_e}$ values) inducing a first-order transition~\footnote{ The QCD EoS applied in this work and derived in \cite{gao2023}  assumes negligible super cooling. Indeed, Ref.~\cite{Gao:2016hks} indicates that the surface tension may be relatively small (compared to $T_c^3$). This means $\Delta p$ remains small and very little vacuum energy is converted into radiation. In such a case there is negligible change to $Y^{\text{ini}}_B, Y^{\text{ini}}_{L_\alpha}$ and $Y^{\text{ini}}_Q$ during the first-order QCD phase transition.}. In this way, we perform a systematic scan and show the resulting first-order region of $(Y^{\text{ini}}_{L_\mu}, Y^{\text{ini}}_{L_\tau})$ as the grey shaded region in Fig.~\ref{fig:pot_genesis}. It turns out that we always have $|\mu_u|\sim |\mu_d|$ such that we expect a simultaneous first-order transition for the $u$ and $d$ quark.
We find that there exists a region around $Y^{\text{ini}}_{L_\mu}=-Y^{\text{ini}}_{L_\tau}$ (c.f.~Fig.~\ref{fig:pot_genesis}) which is capable of inducing a first-order cosmic QCD transition while being consistent with the CMB and BBN bounds without any need for entropy dilution (assuming the stringent limiting case of full-equilibration). However, to obtain successful baryogenesis while inducing a first-order QCD transition, some amount of minimal entropy dilution between the epochs of QCD transition and BBN is needed as discussed in the previous section. A set of such benchmark points of interest is highlighted by the cyan curve in Fig.~\ref{fig:pot_genesis}, which we use to further compute the possibility of observing a GW signal.
Of particular importance for the calculation of the GW spectrum induced by a first-order QCD transition is the critical temperature $T_c$. Due to the so-called focusing effect of QCD \cite{Dore:2022qyz} (the bending of trajectories towards the CEP in its vicinity) the density of trajectories with different $(Y^{\text{ini}}_{L_{\mu}}, Y^{\text{ini}}_{L_{\tau}})$-values in the region around the CEP is expected to be very high.  
Resolving different trajectories in the CEP region would therefore require extremely high numerical precision and in practice this means that we cannot determine the exact value of $T_c$ for given $(Y^{\text{ini}}_{L_{\mu}}, Y^{\text{ini}}_{L_{\tau}})$-values. 
Nevertheless, a conservative estimate for the critical temperatures for the benchmark scenarios in Fig.~\ref{fig:pot_genesis} is provided by the gap in the trajectories, where no solutions can be found. This gives the interval $T_c \in [109,118]$~MeV. 
Note that in principle, going deeper into the region of a first-order QCD transition in Fig.~\ref{fig:pot_genesis} can induce lower values of $T_c$. We here, however, restrict our analysis to the values of the benchmark line as chosen in Fig.~\ref{fig:pot_genesis} because the uncertainty of the determination of $T_c$ increases the deeper we are in the first-order regime. 


\paragraph{Gravitational wave imprint---}

\begin{figure}
    \centering
    \includegraphics[width=\columnwidth]{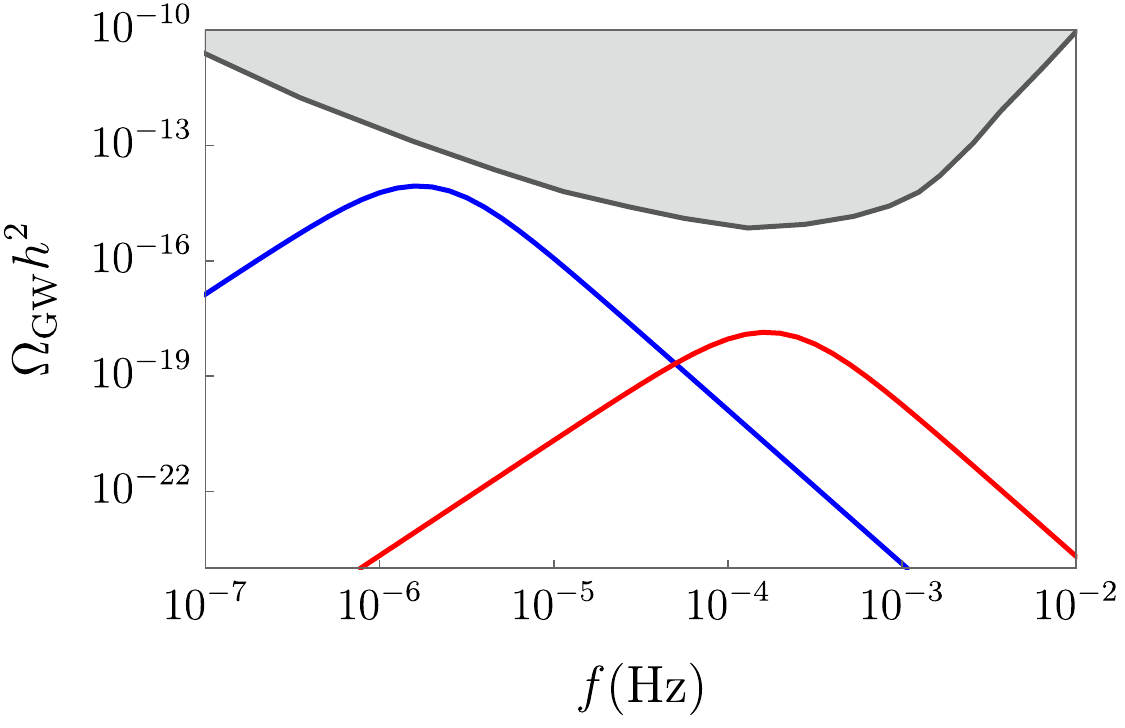}
    \caption{Two gravitational wave spectra with the thermal parameters $(\alpha, v_w, T_c) = (0.1,1,0.109{ \, \rm GeV})$, and the inverse timescales $\beta /H_c = 10^2$ (blue) and $\beta/H_c=10^4$ (orange). The sensitivity of $\mu$Ares is only weakly sensitive to the timescale of the transition for a reasonably typical range.}
    \label{fig:GWbenchmark}
\end{figure}

\begin{figure}
    \centering
    \includegraphics[width=\columnwidth]{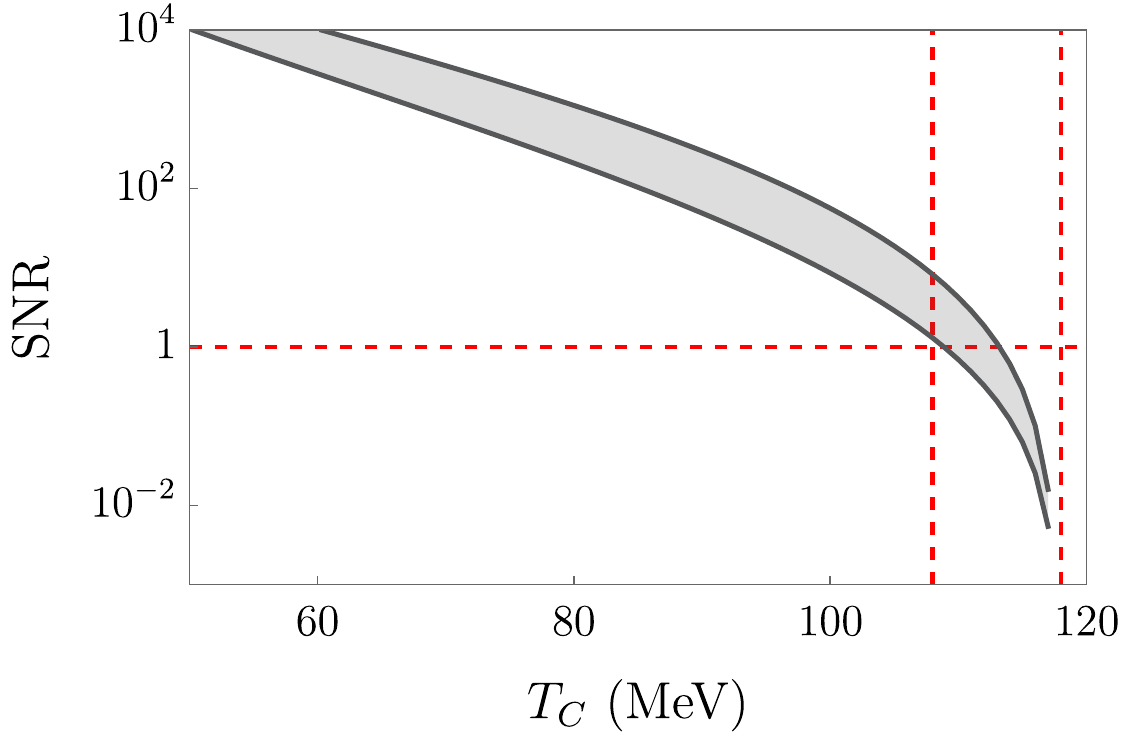}
    \caption{Signal-to-noise ratio as a function of the critical temperature, taking the trace anomaly for two quarks for a phase transition that lasts between $10^{-2}$ and $10^{-4}$ of the Hubble time. The gridlines denote what we expect to be the realistic range of critical temperatures for  the benchmark scenarios marked in cyan in Fig.~\ref{fig:pot_genesis}.}
    \label{fig:SNR}
\end{figure}
Recently, it was proposed to launch three interferometers to the Lagrange points of the Earth's orbit around the sun as a new window into GWs at the $\mu$Hz level. Such a proposal, known as ``$\mu$Ares'' \cite{Sesana:2019vho}, seems perfectly designed to probe the nature of the QCD transition due to its sensitivity to microhertz gravitational waves. In principle if there is a very strong cosmological QCD phase transition, methods sensitive to nanohertz sources such as astrometry or pulsar timing arrays could also detect a signal \cite{Caprini:2010xv,Garcia-Bellido:2021zgu}. Either way, if such a signal manifests, there are very few known scenarios that can mimic the signal from a first-order QCD transition. The two exceptions, a supercooled electroweak phase transition \cite{Kobakhidze:2017mru} and a dark sector transition at the QCD scale \cite{Garcia-Bellido:2021zgu}, can likely be distinguished from a QCD transition by other means (see refs. \cite{Hochberg:2014dra,Ramsey-Musolf:2019lsf} for details), making a GW background from a QCD phase transition a near smoking gun for the sphaleron freeze-in paradigm. 
The GW signal from a strong first-order transition is generally dominated by a contribution arising from sound waves \cite{Caprini:2019egz}.
We calculate the spectrum using the sound shell model \cite{Hindmarsh:2013xza,Hindmarsh:2015qta} in an expanding background \cite{Guo:2020grp,Ellis:2020awk}. In the sound shell model, the GW spectrum grows with the difference in the trace anomaly between the two phases which can be extracted from the EoS of QCD as~\cite{gao2023}:
\begin{equation}
    \theta = \Delta p - T 
\frac{1}{4} \Delta s-\mu\frac{1}{4} \Delta n  \geq \left.\left(- T 
\frac{1}{4} \Delta s-\mu\frac{1}{4} \Delta n\right) \right| _{T_c} .
\end{equation}
Here $T_c$ is the critical temperature at which the pressure difference between the phases vanishes, $(\Delta p,\Delta s, \Delta n)$ are the pressure, entropy and density differences between the phases respectively. We follow the above argument and set the FOPT take place at $\Delta p=0$ (implying percolation temperature $\sim T_c$), and the entropy and density differences  $(\Delta s, \Delta n)$ can be directly read off from the EoS at the two sides of the PT line. For the range of values for $\theta$ that we consider, the signal today has an approximate scaling of $\Omega _{\rm GW}h^2 \sim 10^2 \theta ^3 v_w/\beta ^2$ (for details and exact expressions see \cite{Guo:2020grp}).
Note that the GW signal is monotonic with $\theta$, so saturating the above bound by assuming the transition is at the critical temperature results in a conservative lower bound on the GW peak amplitude. We also assume that the walls are relativistic, noting that the efficiency of the sound-shell source for hybrid and deflagrations is higher unless the bubble wall velocity is very small ($v_w \lesssim 0.3$ \cite{Espinosa:2010hh}). The one remaining parameter is the mean bubble separation, usually denoted as $R_\ast$, or equivalently, the inverse time scale of the transition, usually denoted as $\beta$. This is unfortunately difficult to estimate without knowledge of the surface tension of a bubble of hadronic phase in a medium of a quark-gluon plasma. However, due to the suggested sensitivity profile of $\mu$Ares, the signal-to-noise ratio (SNR) of the expected GW background is not very sensitive to the timescale of the transition. A realistic range for a typical first-order phase transition is between a hundredth and a ten-thousandth of the Hubble time \cite{Caprini:2019egz}, and we plot the SNR for this range as a function of temperature in Fig. \ref{fig:SNR}. The reason for the insensitivity to the transition time can be understood from Fig. \ref{fig:GWbenchmark}. A reduction of the transition time as a fraction of the Hubble time reduces the peak GW amplitude, but also increases the peak frequency, pushing the spectrum closer to the peak sensitivity of 
$\mu$Ares.   

\paragraph{Conclusions---} Using the framework of dimensional reduction at finite temperature and chemical potential we found that the required primordial lepton asymmetry for a successful baryogenesis via sphaleron freeze-in is reduced by an order of magnitude as compared to the previous results from perturbative effective potential. We improved the calculation of the lepton flavor asymmetries that induce a first-order QCD transition in~\cite{Gao:2021nwz} by taking into account QCD thermodynamic quantities with an improved equation of state~\cite{Gao:2023lng}. We find that a first-order QCD phase transition is expected for a substantial part of the baryogenesis parameter space. For a strong first-order QCD transition, the GW spectrum can be detected at $\mu$Ares. This can provide a smoking gun evidence for the sphaleron freeze-in baryogenesis paradigm. A better understanding of the surface tension for a
bubble in the hadronic phase would play a key role in improving the predictions further.   

\begin{acknowledgments}
\paragraph{Acknowledgments---}
The authors thank Travis Dore, Jeff Dror, Hitoshi Murayama, Rob Pisarski, and Alberto Sesana for very useful discussions. FG is supported by the National  Science Foundation of China under Grants  No. 12305134. JH acknowledges support from the Cluster of Excellence “Precision Physics, Fundamental Interactions, and Structure of Matter” (PRISMA$^+$ EXC 2118/1) funded by the Deutsche Forschungsgemeinschaft (DFG, German Research Foundation) within the German Excellence Strategy (Project No. 39083149). CH acknowledges support from the IISN convention No. 4.4503.15, from the CDEIGENT grant No. CIDEIG/2022/16 (funded by Generalitat Valenciana under Plan Gen-T), and from the Spanish grants PID2020-113775GB-I00 (AEI/10.13039/501100011033) and Prometeo CIPROM/2021/054 (Generalitat Valenciana). JH and CH also acknowledge support from the Deutsche Forschungsgemeinschaft (DFG, German Research Foundation) through the Emmy Noether grant HA 8555/1-1. YL is supported by the National  Science Foundation of China under Grants  No. 12175007 and No. 12247107. IMO acknowledges support by Fonds de la recherche scientifique (FRS-FNRS).
\end{acknowledgments}

\bibliographystyle{utphys}
\bibliography{mm}

\providecommand{\href}[2]{#2}\begingroup\raggedright\begin{thebibliography}{10}

\bibitem{NANOGrav:2023gor}
{\bf NANOGrav} Collaboration, G.~Agazie {\em et al.}, ``{The NANOGrav 15 yr
  Data Set: Evidence for a Gravitational-wave Background},''
  \href{http://dx.doi.org/10.3847/2041-8213/acdac6}{{\em Astrophys. J. Lett.}
  {\bf 951} (2023) no.~1, L8}, \href{http://arxiv.org/abs/2306.16213}{{\tt
  arXiv:2306.16213 [astro-ph.HE]}}.

\bibitem{NANOGrav:2023hde}
{\bf NANOGrav} Collaboration, G.~Agazie {\em et al.}, ``{The NANOGrav 15 yr
  Data Set: Observations and Timing of 68 Millisecond Pulsars},''
  \href{http://dx.doi.org/10.3847/2041-8213/acda9a}{{\em Astrophys. J. Lett.}
  {\bf 951} (2023) no.~1, L9}, \href{http://arxiv.org/abs/2306.16217}{{\tt
  arXiv:2306.16217 [astro-ph.HE]}}.

\bibitem{EPTA:2023fyk}
{\bf EPTA} Collaboration, J.~Antoniadis {\em et al.}, ``{The second data
  release from the European Pulsar Timing Array III. Search for gravitational
  wave signals},'' \href{http://arxiv.org/abs/2306.16214}{{\tt arXiv:2306.16214
  [astro-ph.HE]}}.

\bibitem{EPTA:2023sfo}
{\bf EPTA} Collaboration, J.~Antoniadis {\em et al.}, ``{The second data
  release from the European Pulsar Timing Array I. The dataset and timing
  analysis},'' \href{http://arxiv.org/abs/2306.16224}{{\tt arXiv:2306.16224
  [astro-ph.HE]}}.

\bibitem{EPTA:2023xxk}
{\bf EPTA} Collaboration, J.~Antoniadis {\em et al.}, ``{The second data
  release from the European Pulsar Timing Array: V. Implications for massive
  black holes, dark matter and the early Universe},''
  \href{http://arxiv.org/abs/2306.16227}{{\tt arXiv:2306.16227 [astro-ph.CO]}}.

\bibitem{Reardon:2023gzh}
D.~J. Reardon {\em et al.}, ``{Search for an Isotropic Gravitational-wave
  Background with the Parkes Pulsar Timing Array},''
  \href{http://dx.doi.org/10.3847/2041-8213/acdd02}{{\em Astrophys. J. Lett.}
  {\bf 951} (2023) no.~1, L6}, \href{http://arxiv.org/abs/2306.16215}{{\tt
  arXiv:2306.16215 [astro-ph.HE]}}.

\bibitem{Zic:2023gta}
A.~Zic {\em et al.}, ``{The Parkes Pulsar Timing Array Third Data Release},''
  \href{http://arxiv.org/abs/2306.16230}{{\tt arXiv:2306.16230 [astro-ph.HE]}}.

\bibitem{Reardon:2023zen}
D.~J. Reardon {\em et al.}, ``{The Gravitational-wave Background Null
  Hypothesis: Characterizing Noise in Millisecond Pulsar Arrival Times with the
  Parkes Pulsar Timing Array},''
  \href{http://dx.doi.org/10.3847/2041-8213/acdd03}{{\em Astrophys. J. Lett.}
  {\bf 951} (2023) no.~1, L7}, \href{http://arxiv.org/abs/2306.16229}{{\tt
  arXiv:2306.16229 [astro-ph.HE]}}.

\bibitem{Xu:2023wog}
H.~Xu {\em et al.}, ``{Searching for the Nano-Hertz Stochastic Gravitational
  Wave Background with the Chinese Pulsar Timing Array Data Release I},''
  \href{http://dx.doi.org/10.1088/1674-4527/acdfa5}{{\em Res. Astron.
  Astrophys.} {\bf 23} (2023) no.~7, 075024},
  \href{http://arxiv.org/abs/2306.16216}{{\tt arXiv:2306.16216 [astro-ph.HE]}}.

\bibitem{Kuzmin:1985mm}
V.~A. Kuzmin, V.~A. Rubakov, and M.~E. Shaposhnikov, ``{On the Anomalous
  Electroweak Baryon Number Nonconservation in the Early Universe},''
  \href{http://dx.doi.org/10.1016/0370-2693(85)91028-7}{{\em Phys. Lett. B}
  {\bf 155} (1985)  36}.

\bibitem{Davidson:2008bu}
S.~Davidson, E.~Nardi, and Y.~Nir, ``{Leptogenesis},''
  \href{http://dx.doi.org/10.1016/j.physrep.2008.06.002}{{\em Phys. Rept.} {\bf
  466} (2008)  105--177}, \href{http://arxiv.org/abs/0802.2962}{{\tt
  arXiv:0802.2962 [hep-ph]}}.

\bibitem{Chun:2017spz}
E.~J. Chun {\em et al.}, ``{Probing Leptogenesis},''
  \href{http://dx.doi.org/10.1142/S0217751X18420058}{{\em Int. J. Mod. Phys. A}
  {\bf 33} (2018) no.~05n06, 1842005},
  \href{http://arxiv.org/abs/1711.02865}{{\tt arXiv:1711.02865 [hep-ph]}}.

\bibitem{Elor:2022hpa}
G.~Elor {\em et al.}, ``{New Ideas in Baryogenesis: A Snowmass White Paper},''
  in {\em {Snowmass 2021}}.
\newblock 3, 2022.
\newblock \href{http://arxiv.org/abs/2203.05010}{{\tt arXiv:2203.05010
  [hep-ph]}}.

\bibitem{Bajc:1997ky}
B.~Bajc, A.~Riotto, and G.~Senjanovic, ``{Large lepton number of the universe
  and the fate of topological defects},''
  \href{http://dx.doi.org/10.1103/PhysRevLett.81.1355}{{\em Phys. Rev. Lett.}
  {\bf 81} (1998)  1355--1358}, \href{http://arxiv.org/abs/hep-ph/9710415}{{\tt
  arXiv:hep-ph/9710415}}.

\bibitem{McDonald:1999he}
J.~McDonald, ``{Symmetry nonrestoration via order 10**-10 B and L
  asymmetries},'' \href{http://dx.doi.org/10.1016/S0370-2693(99)00992-2}{{\em
  Phys. Lett. B} {\bf 463} (1999)  225--229},
  \href{http://arxiv.org/abs/hep-ph/9907358}{{\tt arXiv:hep-ph/9907358}}.

\bibitem{Bajc:1999he}
B.~Bajc and G.~Senjanovic, ``{Large lepton number and high temperature symmetry
  breaking in MSSM},''
  \href{http://dx.doi.org/10.1016/S0370-2693(99)01432-X}{{\em Phys. Lett. B}
  {\bf 472} (2000)  373--381}, \href{http://arxiv.org/abs/hep-ph/9907552}{{\tt
  arXiv:hep-ph/9907552}}.

\bibitem{McDonald:1999in}
J.~McDonald, ``{Naturally large cosmological neutrino asymmetries in the
  MSSM},'' \href{http://dx.doi.org/10.1103/PhysRevLett.84.4798}{{\em Phys. Rev.
  Lett.} {\bf 84} (2000)  4798--4801},
  \href{http://arxiv.org/abs/hep-ph/9908300}{{\tt arXiv:hep-ph/9908300}}.

\bibitem{March-Russell:1999hpw}
J.~March-Russell, H.~Murayama, and A.~Riotto, ``{The Small observed baryon
  asymmetry from a large lepton asymmetry},''
  \href{http://dx.doi.org/10.1088/1126-6708/1999/11/015}{{\em JHEP} {\bf 11}
  (1999)  015}, \href{http://arxiv.org/abs/hep-ph/9908396}{{\tt
  arXiv:hep-ph/9908396}}.

\bibitem{Barenboim:2017dfq}
G.~Barenboim and W.-I. Park, ``{A full picture of large lepton number
  asymmetries of the Universe},''
  \href{http://dx.doi.org/10.1088/1475-7516/2017/04/048}{{\em JCAP} {\bf 04}
  (2017)  048}, \href{http://arxiv.org/abs/1703.08258}{{\tt arXiv:1703.08258
  [hep-ph]}}.

\bibitem{Schwarz:2009ii}
D.~J. Schwarz and M.~Stuke, ``{Lepton asymmetry and the cosmic QCD
  transition},'' \href{http://dx.doi.org/10.1088/1475-7516/2009/11/025}{{\em
  JCAP} {\bf 11} (2009)  025}, \href{http://arxiv.org/abs/0906.3434}{{\tt
  arXiv:0906.3434 [hep-ph]}}. [Erratum: JCAP 10, E01 (2010)].

\bibitem{Wygas:2018otj}
M.~M. Wygas, I.~M. Oldengott, D.~B\"odeker, and D.~J. Schwarz, ``{Cosmic QCD
  Epoch at Nonvanishing Lepton Asymmetry},''
  \href{http://dx.doi.org/10.1103/PhysRevLett.121.201302}{{\em Phys. Rev.
  Lett.} {\bf 121} (2018) no.~20, 201302},
  \href{http://arxiv.org/abs/1807.10815}{{\tt arXiv:1807.10815 [hep-ph]}}.

\bibitem{Middeldorf-Wygas:2020glx}
M.~M. Middeldorf-Wygas, I.~M. Oldengott, D.~B\"odeker, and D.~J. Schwarz,
  ``{Cosmic QCD transition for large lepton flavor asymmetries},''
  \href{http://dx.doi.org/10.1103/PhysRevD.105.123533}{{\em Phys. Rev. D} {\bf
  105} (2022) no.~12, 123533}, \href{http://arxiv.org/abs/2009.00036}{{\tt
  arXiv:2009.00036 [hep-ph]}}.

\bibitem{Bodeker:2020stj}
D.~Bodeker, F.~Kühnel, I.~M. Oldengott, and D.~J. Schwarz, ``{Lepton Flavour
  Asymmetries and the Mass Spectrum of Primordial Black Holes},''
\href{http://arxiv.org/abs/2011.07283}{{\tt arXiv:2011.07283 [astro-ph.CO]}}.

\bibitem{Vovchenko:2020crk}
V.~Vovchenko, B.~B. Brandt, F.~Cuteri, G.~Endr\H{o}di, F.~Hajkarim, and
  J.~Schaffner-Bielich, ``{Pion Condensation in the Early Universe at
  Nonvanishing Lepton Flavor Asymmetry and Its Gravitational Wave
  Signatures},'' \href{http://dx.doi.org/10.1103/PhysRevLett.126.012701}{{\em
  Phys. Rev. Lett.} {\bf 126} (2021) no.~1, 012701},
  \href{http://arxiv.org/abs/2009.02309}{{\tt arXiv:2009.02309 [hep-ph]}}.

\bibitem{Hajkarim:2019csy}
F.~Hajkarim, J.~Schaffner-Bielich, S.~Wystub, and M.~M. Wygas, ``{Effects of
  the QCD Equation of State and Lepton Asymmetry on Primordial Gravitational
  Waves},'' \href{http://dx.doi.org/10.1103/PhysRevD.99.103527}{{\em Phys. Rev.
  D} {\bf 99} (2019) no.~10, 103527},
  \href{http://arxiv.org/abs/1904.01046}{{\tt arXiv:1904.01046 [hep-ph]}}.

\bibitem{Gao:2021nwz}
F.~Gao and I.~M. Oldengott, ``{Cosmology Meets Functional QCD: First-Order
  Cosmic QCD Transition Induced by Large Lepton Asymmetries},''
  \href{http://dx.doi.org/10.1103/PhysRevLett.128.131301}{{\em Phys. Rev.
  Lett.} {\bf 128} (2022) no.~13, 131301},
  \href{http://arxiv.org/abs/2106.11991}{{\tt arXiv:2106.11991 [hep-ph]}}.

\bibitem{Kajantie:1995dw}
K.~Kajantie, M.~Laine, K.~Rummukainen, and M.~E. Shaposhnikov, ``{Generic rules
  for high temperature dimensional reduction and their application to the
  standard model},'' \href{http://dx.doi.org/10.1016/0550-3213(95)00549-8}{{\em
  Nucl. Phys. B} {\bf 458} (1996)  90--136},
  \href{http://arxiv.org/abs/hep-ph/9508379}{{\tt arXiv:hep-ph/9508379}}.

\bibitem{Braaten:1995cm}
E.~Braaten and A.~Nieto, ``{Effective field theory approach to high temperature
  thermodynamics},'' \href{http://dx.doi.org/10.1103/PhysRevD.51.6990}{{\em
  Phys. Rev. D} {\bf 51} (1995)  6990--7006},
  \href{http://arxiv.org/abs/hep-ph/9501375}{{\tt arXiv:hep-ph/9501375}}.

\bibitem{gao2023}
Y.~Lu, F.~Gao, B.-C. Fu, H.-C. Song, and Y.-X. Liu, ``{Constructing Equation of
  State of QCD in a functional QCD based scheme},'' {\em to appear soon} (2023)
   .

\bibitem{Sesana:2019vho}
A.~Sesana {\em et al.}, ``{Unveiling the gravitational universe at $\mu$-Hz
  frequencies},'' \href{http://dx.doi.org/10.1007/s10686-021-09709-9}{{\em
  Exper. Astron.} {\bf 51} (2021) no.~3, 1333--1383},
  \href{http://arxiv.org/abs/1908.11391}{{\tt arXiv:1908.11391 [astro-ph.IM]}}.

\bibitem{Dolgov:2002ab}
A.~D. Dolgov, S.~H. Hansen, S.~Pastor, S.~T. Petcov, G.~G. Raffelt, and D.~V.
  Semikoz, ``{Cosmological bounds on neutrino degeneracy improved by flavor
  oscillations},'' \href{http://dx.doi.org/10.1016/S0550-3213(02)00274-2}{{\em
  Nucl. Phys. B} {\bf 632} (2002)  363--382},
  \href{http://arxiv.org/abs/hep-ph/0201287}{{\tt arXiv:hep-ph/0201287}}.

\bibitem{Wong:2002fa}
Y.~Y.~Y. Wong, ``{Analytical treatment of neutrino asymmetry equilibration from
  flavor oscillations in the early universe},''
  \href{http://dx.doi.org/10.1103/PhysRevD.66.025015}{{\em Phys. Rev. D} {\bf
  66} (2002)  025015}, \href{http://arxiv.org/abs/hep-ph/0203180}{{\tt
  arXiv:hep-ph/0203180}}.

\bibitem{Oldengott:2017tzj}
I.~M. Oldengott and D.~J. Schwarz, ``{Improved constraints on lepton asymmetry
  from the cosmic microwave background},''
  \href{http://dx.doi.org/10.1209/0295-5075/119/29001}{{\em EPL} {\bf 119}
  (2017) no.~2, 29001}, \href{http://arxiv.org/abs/1706.01705}{{\tt
  arXiv:1706.01705 [astro-ph.CO]}}.

\bibitem{Pitrou:2018cgg}
C.~Pitrou, A.~Coc, J.-P. Uzan, and E.~Vangioni, ``{Precision big bang
  nucleosynthesis with improved Helium-4 predictions},''
  \href{http://dx.doi.org/10.1016/j.physrep.2018.04.005}{{\em Phys. Rept.} {\bf
  754} (2018)  1--66}, \href{http://arxiv.org/abs/1801.08023}{{\tt
  arXiv:1801.08023 [astro-ph.CO]}}.

\bibitem{Note1}
Interestingly, recent measurements of the primordial helium abundance by the
  EMPRESS survey \cite {Matsumoto:2022tlr} could be interpreted as a hint
  towards a positive asymmetry in electron neutrinos \cite
  {Escudero:2022okz,Matsumoto:2022tlr}.

\bibitem{Pastor:2008ti}
S.~Pastor, T.~Pinto, and G.~G. Raffelt, ``{Relic density of neutrinos with
  primordial asymmetries},''
  \href{http://dx.doi.org/10.1103/PhysRevLett.102.241302}{{\em Phys. Rev.
  Lett.} {\bf 102} (2009)  241302}, \href{http://arxiv.org/abs/0808.3137}{{\tt
  arXiv:0808.3137 [astro-ph]}}.

\bibitem{Mangano:2010ei}
G.~Mangano, G.~Miele, S.~Pastor, O.~Pisanti, and S.~Sarikas, ``{Constraining
  the cosmic radiation density due to lepton number with Big Bang
  Nucleosynthesis},''
  \href{http://dx.doi.org/10.1088/1475-7516/2011/03/035}{{\em JCAP} {\bf 03}
  (2011)  035}, \href{http://arxiv.org/abs/1011.0916}{{\tt arXiv:1011.0916
  [astro-ph.CO]}}.

\bibitem{Mangano:2011ip}
G.~Mangano, G.~Miele, S.~Pastor, O.~Pisanti, and S.~Sarikas, ``{Updated BBN
  bounds on the cosmological lepton asymmetry for non-zero $\theta_{13}$},''
  \href{http://dx.doi.org/10.1016/j.physletb.2012.01.015}{{\em Phys. Lett. B}
  {\bf 708} (2012)  1--5}, \href{http://arxiv.org/abs/1110.4335}{{\tt
  arXiv:1110.4335 [hep-ph]}}.

\bibitem{Castorina:2012md}
E.~Castorina, U.~Franca, M.~Lattanzi, J.~Lesgourgues, G.~Mangano,
  A.~Melchiorri, and S.~Pastor, ``{Cosmological lepton asymmetry with a nonzero
  mixing angle $\theta_{13}$},''
  \href{http://dx.doi.org/10.1103/PhysRevD.86.023517}{{\em Phys. Rev. D} {\bf
  86} (2012)  023517}, \href{http://arxiv.org/abs/1204.2510}{{\tt
  arXiv:1204.2510 [astro-ph.CO]}}.

\bibitem{Barenboim:2016shh}
G.~Barenboim, W.~H. Kinney, and W.-I. Park, ``{Resurrection of large lepton
  number asymmetries from neutrino flavor oscillations},''
  \href{http://dx.doi.org/10.1103/PhysRevD.95.043506}{{\em Phys. Rev. D} {\bf
  95} (2017) no.~4, 043506}, \href{http://arxiv.org/abs/1609.01584}{{\tt
  arXiv:1609.01584 [hep-ph]}}.

\bibitem{Johns:2016enc}
L.~Johns, M.~Mina, V.~Cirigliano, M.~W. Paris, and G.~M. Fuller, ``{Neutrino
  flavor transformation in the lepton-asymmetric universe},''
  \href{http://dx.doi.org/10.1103/PhysRevD.94.083505}{{\em Phys. Rev.} {\bf
  D94} (2016) no.~8, 083505},
\href{http://arxiv.org/abs/1608.01336}{{\tt arXiv:1608.01336 [hep-ph]}}.

\bibitem{Domcke:2022uue}
V.~Domcke, K.~Kamada, K.~Mukaida, K.~Schmitz, and M.~Yamada, ``{A new
  constraint on primordial lepton flavour asymmetries},''
  \href{http://arxiv.org/abs/2208.03237}{{\tt arXiv:2208.03237 [hep-ph]}}.

\bibitem{Joyce:1997uy}
M.~Joyce and M.~E. Shaposhnikov, ``{Primordial magnetic fields, right-handed
  electrons, and the Abelian anomaly},''
  \href{http://dx.doi.org/10.1103/PhysRevLett.79.1193}{{\em Phys. Rev. Lett.}
  {\bf 79} (1997)  1193--1196},
  \href{http://arxiv.org/abs/astro-ph/9703005}{{\tt arXiv:astro-ph/9703005}}.

\bibitem{Linde:1980ts}
A.~D. Linde, ``{Infrared Problem in Thermodynamics of the Yang-Mills Gas},''
  \href{http://dx.doi.org/10.1016/0370-2693(80)90769-8}{{\em Phys. Lett. B}
  {\bf 96} (1980)  289--292}.

\bibitem{Gould:2021oba}
O.~Gould and T.~V.~I. Tenkanen, ``{On the perturbative expansion at high
  temperature and implications for cosmological phase transitions},''
  \href{http://dx.doi.org/10.1007/JHEP06(2021)069}{{\em JHEP} {\bf 06} (2021)
  069}, \href{http://arxiv.org/abs/2104.04399}{{\tt arXiv:2104.04399
  [hep-ph]}}.

\bibitem{Curtin:2022ovx}
D.~Curtin, J.~Roy, and G.~White, ``{Gravitational waves and tadpole
  resummation: Efficient and easy convergence of finite temperature QFT},''
  \href{http://arxiv.org/abs/2211.08218}{{\tt arXiv:2211.08218 [hep-ph]}}.

\bibitem{Croon:2020cgk}
D.~Croon, O.~Gould, P.~Schicho, T.~V.~I. Tenkanen, and G.~White, ``{Theoretical
  uncertainties for cosmological first-order phase transitions},''
  \href{http://dx.doi.org/10.1007/JHEP04(2021)055}{{\em JHEP} {\bf 04} (2021)
  055}, \href{http://arxiv.org/abs/2009.10080}{{\tt arXiv:2009.10080
  [hep-ph]}}.

\bibitem{Gynther:2003za}
A.~Gynther, ``{Electroweak phase diagram at finite lepton number density},''
  \href{http://dx.doi.org/10.1103/PhysRevD.68.016001}{{\em Phys. Rev. D} {\bf
  68} (2003)  016001}, \href{http://arxiv.org/abs/hep-ph/0303019}{{\tt
  arXiv:hep-ph/0303019}}.

\bibitem{Gao:2023lng}
F.~Gao, J.~Harz, C.~Hati, I.~M. Oldengott, Y.~Lu, and G.~White, ``{To appear
  soon},''.

\bibitem{Carson:1990jm}
L.~Carson, X.~Li, L.~D. McLerran, and R.-T. Wang, ``{Exact Computation of the
  Small Fluctuation Determinant Around a Sphaleron},''
  \href{http://dx.doi.org/10.1103/PhysRevD.42.2127}{{\em Phys. Rev. D} {\bf 42}
  (1990)  2127--2143}.

\bibitem{Note2}
An amount of $Y_{L_e}$ will be generated once the neutrino oscillations are in
  equilibrium shortly before BBN, however, it has no impact on our analysis.

\bibitem{Moroi:1999zb}
T.~Moroi and L.~Randall, ``{Wino cold dark matter from anomaly mediated SUSY
  breaking},'' \href{http://dx.doi.org/10.1016/S0550-3213(99)00748-8}{{\em
  Nucl. Phys. B} {\bf 570} (2000)  455--472},
  \href{http://arxiv.org/abs/hep-ph/9906527}{{\tt arXiv:hep-ph/9906527}}.

\bibitem{Thomas:1995ze}
S.~D. Thomas, ``{Baryons and dark matter from the late decay of a
  supersymmetric condensate},''
  \href{http://dx.doi.org/10.1016/0370-2693(95)00772-D}{{\em Phys. Lett. B}
  {\bf 356} (1995)  256--263}, \href{http://arxiv.org/abs/hep-ph/9506274}{{\tt
  arXiv:hep-ph/9506274}}.

\bibitem{Moroi:1994rs}
T.~Moroi, M.~Yamaguchi, and T.~Yanagida, ``{On the solution to the Polonyi
  problem with 0 (10-TeV) gravitino mass in supergravity},''
  \href{http://dx.doi.org/10.1016/0370-2693(94)01337-C}{{\em Phys. Lett. B}
  {\bf 342} (1995)  105--110}, \href{http://arxiv.org/abs/hep-ph/9409367}{{\tt
  arXiv:hep-ph/9409367}}.

\bibitem{Allahverdi:2002nb}
R.~Allahverdi and M.~Drees, ``{Production of massive stable particles in
  inflaton decay},''
  \href{http://dx.doi.org/10.1103/PhysRevLett.89.091302}{{\em Phys. Rev. Lett.}
  {\bf 89} (2002)  091302}, \href{http://arxiv.org/abs/hep-ph/0203118}{{\tt
  arXiv:hep-ph/0203118}}.

\bibitem{Moroi:2002rd}
T.~Moroi and T.~Takahashi, ``{Cosmic density perturbations from late decaying
  scalar condensations},''
  \href{http://dx.doi.org/10.1103/PhysRevD.66.063501}{{\em Phys. Rev. D} {\bf
  66} (2002)  063501}, \href{http://arxiv.org/abs/hep-ph/0206026}{{\tt
  arXiv:hep-ph/0206026}}.

\bibitem{Lahanas:2011tk}
A.~B. Lahanas, ``{Dilaton dominance in the early Universe dilutes Dark Matter
  relic abundances},'' \href{http://dx.doi.org/10.1103/PhysRevD.83.103523}{{\em
  Phys. Rev. D} {\bf 83} (2011)  103523},
  \href{http://arxiv.org/abs/1102.4277}{{\tt arXiv:1102.4277 [hep-ph]}}.

\bibitem{Fujii:2002kr}
M.~Fujii and K.~Hamaguchi, ``{Nonthermal dark matter via Affleck-Dine
  baryogenesis and its detection possibility},''
  \href{http://dx.doi.org/10.1103/PhysRevD.66.083501}{{\em Phys. Rev. D} {\bf
  66} (2002)  083501}, \href{http://arxiv.org/abs/hep-ph/0205044}{{\tt
  arXiv:hep-ph/0205044}}.

\bibitem{Gao:2015kea}
F.~Gao, J.~Chen, Y.-X. Liu, S.-X. Qin, C.~D. Roberts, and S.~M. Schmidt,
  ``{Phase diagram and thermal properties of strong-interaction matter},''
  \href{http://dx.doi.org/10.1103/PhysRevD.93.094019}{{\em Phys. Rev. D} {\bf
  93} (2016) no.~9, 094019}, \href{http://arxiv.org/abs/1507.00875}{{\tt
  arXiv:1507.00875 [nucl-th]}}.

\bibitem{Gao:2020qsj}
F.~Gao and J.~M. Pawlowski, ``{QCD phase structure from functional methods},''
  \href{http://dx.doi.org/10.1103/PhysRevD.102.034027}{{\em Phys. Rev. D} {\bf
  102} (2020) no.~3, 034027}, \href{http://arxiv.org/abs/2002.07500}{{\tt
  arXiv:2002.07500 [hep-ph]}}.

\bibitem{Gao:2020fbl}
F.~Gao and J.~M. Pawlowski, ``{Chiral phase structure and critical end point in
  QCD},'' \href{http://dx.doi.org/10.1016/j.physletb.2021.136584}{{\em Phys.
  Lett. B} {\bf 820} (2021)  136584},
  \href{http://arxiv.org/abs/2010.13705}{{\tt arXiv:2010.13705 [hep-ph]}}.

\bibitem{Note3}
The QCD EoS applied in this work and derived in \cite {gao2023} assumes
  negligible super cooling. Indeed, Ref.~\cite {Gao:2016hks} indicates that the
  surface tension may be relatively small (compared to $T_c^3$). This means
  $\Delta p$ remains small and very little vacuum energy is converted into
  radiation. In such a case there is negligible change to $Y^{\protect \text
  {ini}}_B, Y^{\protect \text {ini}}_{L_\alpha }$ and $Y^{\protect \text
  {ini}}_Q$ during the first-order QCD phase transition.

\bibitem{Dore:2022qyz}
T.~Dore, J.~M. Karthein, I.~Long, D.~Mroczek, J.~Noronha-Hostler, P.~Parotto,
  C.~Ratti, and Y.~Yamauchi, ``{Critical lensing and kurtosis near a critical
  point in the QCD phase diagram in and out of equilibrium},''
  \href{http://dx.doi.org/10.1103/PhysRevD.106.094024}{{\em Phys. Rev. D} {\bf
  106} (2022) no.~9, 094024}, \href{http://arxiv.org/abs/2207.04086}{{\tt
  arXiv:2207.04086 [nucl-th]}}.

\bibitem{Caprini:2010xv}
C.~Caprini, R.~Durrer, and X.~Siemens, ``{Detection of gravitational waves from
  the QCD phase transition with pulsar timing arrays},''
  \href{http://dx.doi.org/10.1103/PhysRevD.82.063511}{{\em Phys. Rev. D} {\bf
  82} (2010)  063511}, \href{http://arxiv.org/abs/1007.1218}{{\tt
  arXiv:1007.1218 [astro-ph.CO]}}.

\bibitem{Garcia-Bellido:2021zgu}
J.~Garcia-Bellido, H.~Murayama, and G.~White, ``{Exploring the early Universe
  with Gaia and Theia},''
  \href{http://dx.doi.org/10.1088/1475-7516/2021/12/023}{{\em JCAP} {\bf 12}
  (2021) no.~12, 023}, \href{http://arxiv.org/abs/2104.04778}{{\tt
  arXiv:2104.04778 [hep-ph]}}.

\bibitem{Kobakhidze:2017mru}
A.~Kobakhidze, C.~Lagger, A.~Manning, and J.~Yue, ``{Gravitational waves from a
  supercooled electroweak phase transition and their detection with pulsar
  timing arrays},''
  \href{http://dx.doi.org/10.1140/epjc/s10052-017-5132-y}{{\em Eur. Phys. J. C}
  {\bf 77} (2017) no.~8, 570}, \href{http://arxiv.org/abs/1703.06552}{{\tt
  arXiv:1703.06552 [hep-ph]}}.

\bibitem{Hochberg:2014dra}
Y.~Hochberg, E.~Kuflik, T.~Volansky, and J.~G. Wacker, ``{Mechanism for Thermal
  Relic Dark Matter of Strongly Interacting Massive Particles},''
  \href{http://dx.doi.org/10.1103/PhysRevLett.113.171301}{{\em Phys. Rev.
  Lett.} {\bf 113} (2014)  171301}, \href{http://arxiv.org/abs/1402.5143}{{\tt
  arXiv:1402.5143 [hep-ph]}}.

\bibitem{Ramsey-Musolf:2019lsf}
M.~J. Ramsey-Musolf, ``{The electroweak phase transition: a collider target},''
  \href{http://dx.doi.org/10.1007/JHEP09(2020)179}{{\em JHEP} {\bf 09} (2020)
  179}, \href{http://arxiv.org/abs/1912.07189}{{\tt arXiv:1912.07189
  [hep-ph]}}.

\bibitem{Caprini:2019egz}
C.~Caprini {\em et al.}, ``{Detecting gravitational waves from cosmological
  phase transitions with LISA: an update},''
  \href{http://dx.doi.org/10.1088/1475-7516/2020/03/024}{{\em JCAP} {\bf 03}
  (2020)  024}, \href{http://arxiv.org/abs/1910.13125}{{\tt arXiv:1910.13125
  [astro-ph.CO]}}.

\bibitem{Hindmarsh:2013xza}
M.~Hindmarsh, S.~J. Huber, K.~Rummukainen, and D.~J. Weir, ``{Gravitational
  waves from the sound of a first order phase transition},''
  \href{http://dx.doi.org/10.1103/PhysRevLett.112.041301}{{\em Phys. Rev.
  Lett.} {\bf 112} (2014)  041301}, \href{http://arxiv.org/abs/1304.2433}{{\tt
  arXiv:1304.2433 [hep-ph]}}.

\bibitem{Hindmarsh:2015qta}
M.~Hindmarsh, S.~J. Huber, K.~Rummukainen, and D.~J. Weir, ``{Numerical
  simulations of acoustically generated gravitational waves at a first order
  phase transition},'' \href{http://dx.doi.org/10.1103/PhysRevD.92.123009}{{\em
  Phys. Rev. D} {\bf 92} (2015) no.~12, 123009},
  \href{http://arxiv.org/abs/1504.03291}{{\tt arXiv:1504.03291 [astro-ph.CO]}}.

\bibitem{Guo:2020grp}
H.-K. Guo, K.~Sinha, D.~Vagie, and G.~White, ``{Phase Transitions in an
  Expanding Universe: Stochastic Gravitational Waves in Standard and
  Non-Standard Histories},''
  \href{http://dx.doi.org/10.1088/1475-7516/2021/01/001}{{\em JCAP} {\bf 01}
  (2021)  001}, \href{http://arxiv.org/abs/2007.08537}{{\tt arXiv:2007.08537
  [hep-ph]}}.

\bibitem{Ellis:2020awk}
J.~Ellis, M.~Lewicki, and J.~M. No, ``{Gravitational waves from first-order
  cosmological phase transitions: lifetime of the sound wave source},''
  \href{http://dx.doi.org/10.1088/1475-7516/2020/07/050}{{\em JCAP} {\bf 07}
  (2020)  050}, \href{http://arxiv.org/abs/2003.07360}{{\tt arXiv:2003.07360
  [hep-ph]}}.

\bibitem{Espinosa:2010hh}
J.~R. Espinosa, T.~Konstandin, J.~M. No, and G.~Servant, ``{Energy Budget of
  Cosmological First-order Phase Transitions},''
  \href{http://dx.doi.org/10.1088/1475-7516/2010/06/028}{{\em JCAP} {\bf 06}
  (2010)  028}, \href{http://arxiv.org/abs/1004.4187}{{\tt arXiv:1004.4187
  [hep-ph]}}.

\bibitem{Matsumoto:2022tlr}
A.~Matsumoto {\em et al.}, ``{EMPRESS. VIII. A New Determination of Primordial
  He Abundance with Extremely Metal-poor Galaxies: A Suggestion of the Lepton
  Asymmetry and Implications for the Hubble Tension},''
  \href{http://dx.doi.org/10.3847/1538-4357/ac9ea1}{{\em Astrophys. J.} {\bf
  941} (2022) no.~2, 167}, \href{http://arxiv.org/abs/2203.09617}{{\tt
  arXiv:2203.09617 [astro-ph.CO]}}.

\bibitem{Escudero:2022okz}
M.~Escudero, A.~Ibarra, and V.~Maura, ``{Primordial Lepton Asymmetries in the
  Precision Cosmology Era: Current Status and Future Sensitivities from BBN and
  the CMB},'' \href{http://arxiv.org/abs/2208.03201}{{\tt arXiv:2208.03201
  [hep-ph]}}.

\bibitem{Gao:2016hks}
F.~Gao and Y.-x. Liu, ``{Interface Effect in QCD Phase Transitions via
  Dyson-Schwinger Equation Approach},''
  \href{http://dx.doi.org/10.1103/PhysRevD.94.094030}{{\em Phys. Rev. D} {\bf
  94} (2016) no.~9, 094030}, \href{http://arxiv.org/abs/1609.08038}{{\tt
  arXiv:1609.08038 [hep-ph]}}.

\end{thebibliography}\endgroup

\end{document}